\documentclass[prl,twocolumn]{revtex4-2}

\usepackage[T1]{fontenc}
\usepackage[utf8]{inputenc}
\usepackage{lmodern}
\usepackage{amsfonts}
\usepackage{amsmath}
\usepackage{amssymb}
\usepackage{graphicx}
\usepackage{hyperref}
\usepackage{siunitx}

\hypersetup{
  colorlinks=true,
  linkcolor=blue,
  citecolor=blue,
  urlcolor=blue
}

\begin{document}

\title[aMChA quartz]{Observation of acoustic magneto-chiral anisotropy in $\alpha$-quartz}

\author{Munkhtuguldur Altangerel}
\author{S.~Badoux}
\author{C.~Proust}
\author{D.~Vignolles}
\author{G.~L.~J.~A.~Rikken}
\affiliation{Univ. Toulouse, INSA-T, Univ. Grenoble Alpes, CNRS, LNCMI-EMFL, UPR3228, Toulouse, France.}

\date{\today}

\begin{abstract}
We report the experimental observation of magneto-chiral anisotropy in the longitudinal and transverse ultrasound propagation in $\alpha$-quartz. To perform such measurement, we have built an ultrasound spectrometer with unprecedented experimental resolution of the order of $\Delta v/v \sim 10^{-8}$.  We present a simple macroscopic Becquerel-like analytical model that accounts for the magnitude of the observed effect and its frequency dependence.
\end{abstract}

\maketitle

Chirality plays an important role in many areas of physics, chemistry and
biology, where entities exist in two non-superimposable forms (enantiomers),
one being the mirror image of the other. Chirality corresponds to an absence
of mirror symmetry and if time-reversal symmetry is also absent because of a
magnetization or an external magnetic field, an entire class of effects
called magneto-chiral anisotropy (MChA) becomes allowed. Optical MChA
corresponds to a difference in the absorption and refraction of unpolarized
light propagating through the chiral medium parallel or anti-parallel to the
field \cite{WagniereMeier,BarronVbrancich}. Initially observed in the
visible wavelength range \cite{Naturemca,kleindienst,mcaabs}, its existence
was later confirmed across the entire electromagnetic spectrum, from
microwaves \cite{MicrowaveMChA} to X-rays \cite{XrayMChA}. MChA was further
generalized to other transport phenomena \cite{emchaprl}. It was
experimentally observed in the electrical transport in bismuth helices \cite%
{emchaprl}, in carbon nanotubes \cite{eMChACNT}, in bulk organic conductors 
\cite{Pop}, in metals \cite{Yokouchi,Aoki}, in superconductors close to the
transition temperature \cite{MChAsupercon} and in semiconductors \cite%
{RikkenTe} as an electrical resistance that depends on the handedness of the
conductor and on the relative orientation of electrical current and magnetic
field. Other manifestations of MChA were observed in the propagation of
magnons \cite{MagnonMChA} and of ultrasound \cite{USMChA1,USMChA2} in chiral
magnetic crystals, and in the displacement current in chiral dielectrics 
\cite{dMChA-Rikken}, further illustrating its universal character. Inverse
MChA, a longitudinal magnetization induced in a chiral medium by an
unpolarized flux, first proposed in the optical domain \cite{Theory-ioMChA},
and later for an electrical current \cite{Theory-ieMChA}, has also been
observed \cite{Exp-ieMChA}. MChA has become a prominent representative of
the wider class of non-reciprocal transport phenomena in broken-symmetry
systems, that play an important role in topological quantum systems and in
Berry phase physics \cite{Non-reciproc-review}. For a recent review of MChA,
see \cite{Review-MChA}.

The existence of MChA in phonon transport was already predicted a long time
ago \cite{emchaprl}, but it was only more recently in magnetic crystals with
a hybridization between magnons and phonons, that the effect could be
observed as non-reciprocal ultrasound transmission close to a magnetic phase
transition \cite{USMChA1,USMChA2}. These observations are therefore better
qualified as phonon-detected MagnonMChA. Although the magnitude of the
effect was too small to be useful in phononics, this effect represents an
acoustical diode that can be inverted with a magnetic field, a highly
interesting functionality. It is therefore worthwhile to explore MChA in
phonon transport in other materials. Here we will extend the MChA family to
ultrasound propagation in diamagnetic crystals, in particular in the
archetypal chiral crystal of $\alpha$-quartz.

For optical MChA (oMChA), a detailed quantitative microscopic theory has
been developed \cite{BarronVbrancich} and experimentally validated for
transition-metal complexes \cite{ValidationMChA}. A specific theory for
phonon MChA was developed for chiral Weyl semimetals \cite{Phonon-MChA-Weyl}%
, but so far, no general prediction for diamagnetic crystals exists. For
simplicity we will consider for the remainder only the case of isotropic
media, or uniaxial media aligned with the magnetic field. Symmetry arguments
tell us that to first order the acoustic wave velocity $v$ parallel to the
magnetic field for transverse circularly polarized modes ($\pm$) and left ($L
$) or right ($D$) handed media is given by 
\begin{equation}
v_{\pm }(\omega,\mathbf{k},\mathbf{B})=v(\omega)\pm \alpha^{D/L}(\omega)k\pm
\beta(\omega)B+\gamma^{D/L}(\omega)\,\mathbf{k}\!\cdot\!\mathbf{B},
\label{sym2}
\end{equation}
with $X^{D}=-X^{L}$, where $\alpha$ describes acoustic activity, $\beta$
describes the acoustic Faraday effect and $\gamma$ describes the \textit{%
polarization-independent} acoustic magneto-chiral anisotropy (aMChA). For
longitudinal modes, the $\alpha$ and $\beta$ terms are non-existent, but
this is not the case for the $\gamma$ term. Note that this argument does not
explicitly invoke chiral phonons, but the characteristic of $\alpha\neq 0$
results from the existence of non-degenerate chiral transverse phonons. The
symmetry argument does not give an indication of the strength for aMChA. For
that, we will estimate a value for $\gamma$ with a simple macroscopic
Becquerel-type model \cite{Sup-mat}: In this model, the apparent frequency
that the moving charges experience because of the magnetic field is found to
be shifted away from the externally imposed frequency by the Larmor
frequency $\Omega_{L}=qB/(2m)$, where $q$ is the charge and $m$ the mass of
the charge, the direction of the shift depending on whether the cyclotron
rotation has the same sense as the circular polarization of the incident
wave, or the opposite one. We will therefore approximate the effect of the
magnetic field on the transverse wave velocity in a chiral medium as a
frequency shift by an effective Larmor frequency $\widetilde{\Omega}%
_{L}\equiv \widetilde{q}/(2\widetilde{m})$. This leads to \cite{Sup-mat} 
\begin{widetext}
\begin{equation}
v_{\pm}^{D/L}(\omega,k,B) \simeq v(\omega \pm \tilde{\Omega}_L)
 \pm \alpha^{D/L}(\omega \pm \tilde{\Omega}_L) k
 \simeq v(\omega)
 \pm \alpha^{D/L}(\omega) k
 + \frac{1}{2}\frac{\tilde{q}}{\tilde{m}} \frac{\partial v}{\partial \omega} B
 + \frac{1}{2}\frac{\tilde{q}}{\tilde{m}} \frac{\partial \alpha^{D/L}}{\partial \omega} B k.
 \label{Chiral Becquerel}
\end{equation}
\end{widetext}

where the third term on the right-hand side describes the acoustic Faraday
effect and the fourth term describes the aMChA. Comparing Eq.~\eqref{sym2}
with Eq.~\eqref{Chiral Becquerel} provides estimates for the $\beta$ and $%
\gamma$ parameters in Eq.~\eqref{sym2}, where $\widetilde{q}$ and $%
\widetilde{m}$ are effective values representing the moving charges.

Acoustic activity theory usually limits itself to first-order spatial
dispersion, yielding an $\alpha$ independent of the wave vector \cite%
{Portigal-Burstein}. This agrees well with most experimental results for
small wave vectors \cite{Shen}, but both theory and experiment acknowledge
the existence of second-order spatial dispersion contributions, i.e. $%
\alpha(\omega)=\alpha_{0}+\sigma k$ \cite{Shen}. The
polarization-independent aMChA can therefore be expressed as 
\begin{equation}
\frac{\Delta v}{v_{0}}\equiv \frac{v(B)-v(-B)}{v_{0}}=\frac{\widetilde{q}%
\,\sigma\,\omega\, B}{\widetilde{m}\,v_{0}^{3}}.  \label{deltav/v}
\end{equation}
Note that the predicted linear frequency dependence of aMChA for this case
is different from the cubic frequency dependence predicted and observed for
the magnon-phonon hybridization case \cite{USMChA1,USMChA2}. No explicit
theory for $\sigma$ has been put forward so far, but its value can be
empirically determined from literature values of the acoustic activity of $%
\alpha$-quartz at different frequencies, as summarized below:

\begin{center}
\begin{tabular}{|c|c|c|}
\hline
Frequency (GHz) & $\Delta v/(v f)\times 10^{-13}\,\mathrm{Hz}^{-1}$ & 
Reference \\ \hline
1.4 & 2.7 & \cite{Pine} \\ 
9.4 & 3.0 & \cite{Joffrin} \\ 
9.4 & 2.8 & \cite{Bialas} \\ 
28.9 & 3.3 & \cite{Pine} \\ \hline
\end{tabular}
\end{center}

From this, we find $\sigma = (2.5 \pm 1)\times 10^{-15}\,\mathrm{%
m^{3}\,s^{-1}}$ \cite{Sup-mat}.

In contrast to oMChA, in the case of aMChA, the magnetic field acts, through
the Lorentz force, both on the nuclear movement and on the bound electrons,
and both act upon each other. It is therefore not obvious what value for $%
\widetilde{q}/\widetilde{m}$ to use in Eq.~\eqref{deltav/v}. It will be some
effective value between that of an electron ($1.7\times 10^{11}\,\mathrm{%
C\,kg^{-1}}$) and that of nuclei ($\sim 5\times 10^{7}\,\mathrm{C\,kg^{-1}}$%
). It does seem very plausible that it will be the same value as in the
acoustic Faraday effect, as described by the third term in Eq.~\eqref{Chiral
Becquerel}. This value can be estimated by the following argument: In the
absence of chirality and magnetic field, the phonon dispersion relation is
given by 
\begin{equation}
\omega = v_{0}k - \zeta k^{3}
\end{equation}
and therefore

\begin{equation}
\frac{\partial v}{\partial \omega }=-\frac{6\zeta \omega }{v_{0}^{2}}.
\label{dvdw}
\end{equation}%
Comparing Eq.~\eqref{sym2} with Eq.~\eqref{Chiral Becquerel} and expressing $%
\beta $ in the acoustic Verdet constant $V$ \cite{Sup-mat} gives 
\begin{equation}
\beta =\frac{1}{2}\frac{\widetilde{q}}{\widetilde{m}}\frac{\partial v}{%
\partial \omega }=-\frac{v_{0}^{2}V}{2\omega },  \label{beta as dvdw}
\end{equation}%
and upon combining Eqs.~\eqref{dvdw} and \eqref{beta as dvdw} we obtain 
\begin{equation}
\frac{\widetilde{q}}{\widetilde{m}}=\frac{Vv_{0}^{4}}{6\zeta \omega ^{2}}.
\end{equation}

Using the experimental value for $V$ at \SI{254}{MHz}, $(32\pm3)%
\times10^{-3}~\mathrm{rad\ T^{-1} m^{-1}}$ \cite{Sup-mat}, and the
experimental literature value for $\zeta$, $(2.1\pm0.2)\times 10^{-16}\,%
\mathrm{m^{3}\,s^{-1}}$ \cite{Joffrin1980}, we find $\widetilde{q}/%
\widetilde{m}=(4.1\pm0.6)\times 10^{9}\,\mathrm{C\,kg^{-1}}$. This value,
intermediate between the nuclear and the electronic limits, but closer to
the latter, reveals that there is indeed a significant electronic
contribution to the acoustic Faraday effect in $\alpha$-quartz. With Eq.~%
\eqref{deltav/v} this yields $\Delta v/v_{0}\sim1\times 10^{-6}$ at \SI{1}{T}
and \SI{1}{GHz}. Note that this value is similar to the one predicted for
chiral Weyl semimetals \cite{Phonon-MChA-Weyl}.

So far our argument has been limited to transverse phonons. For atoms
arranged on a helicoidal structure like in $\alpha$-quartz, the transverse
and longitudinal movements of atoms are coupled. Indeed, detailed phonon
dispersion relation calculations \cite{Chiral-phonon-database} show that
longitudinal modes in chiral crystals have some helicity and are therefore
chiral and no longer purely longitudinal. So also for longitudinal phonons,
the cyclotron motion will couple to the lattice vibrations. Also, the theory
for chiral Weyl semimetals predicts aMChA for longitudinal phonons \cite%
{Phonon-MChA-Weyl}. Although not strictly justified, we will assume that Eq.~%
\eqref{deltav/v} is also valid for longitudinal phonons, where $\zeta =
(7.75\pm0.05)\times 10^{-17}\,\mathrm{m^{3}\,s^{-1}}$ \cite{Hao}. Eq.~%
\eqref{deltav/v}, using the same effective value for $\widetilde{q}/%
\widetilde{m}$, predicts for this case $\Delta v/v_{0}\sim2\times 10^{-7}$
at \SI{1}{T} and \SI{1}{GHz}.

To observe such small non-reciprocities, we have designed and constructed a
dedicated ultrasound setup, as schematically shown in Fig.~\ref{fig:schema} 
and described in detail in the Supplementary Information. The
improved sensitivity of this setup as compared to existing techniques
results from (i) the use of an interferometer, which doubles the aMChA
response whereas at the same time it suppresses common mode artefacts (ii) the use of an alternating magnetic field and phase sensitive detection of
the interferometer output on the magnetic field frequency.
\begin{figure}[t]
\centering  \includegraphics[width=0.8\linewidth]{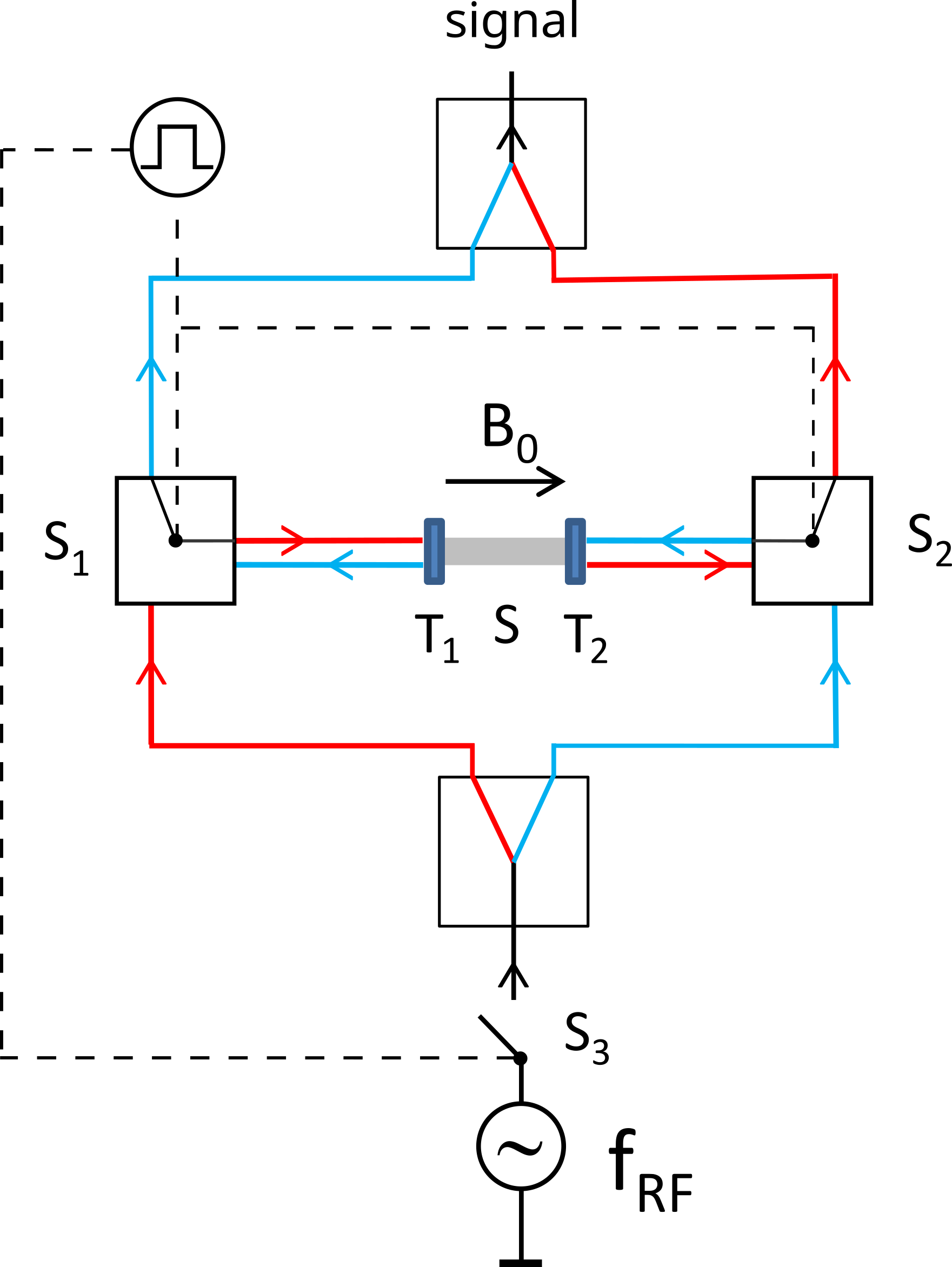}  
\caption{Schematic view of the ultrasound interferometer. RF switches S1, S2
and S3 direct RF excitation pulses towards the ZnO transducers T1 and T2, or
the transmitted signals from the transducers to a differential amplifier
(the situation shown).}
\label{fig:schema}
\end{figure}

Transverse or longitudinal short ultrasound pulses with mW power are
simultaneously injected with piezoelectric ZnO transducers sputtered on both
ends of  $z$-cut $\alpha $-quartz crystals (typical length \SI{10}{mm})
aligned parallel to the external magnetic field. The difference between the
two transmitted counterpropagating pulses, proportional to the
non-reciprocity, is then down-converted and the resulting signal is
phase-sensitively detected. This results in a typical experimental
resolution of $\Delta v/v_{0}\sim 10^{-8}$ at \SI{500}{MHz} and \SI{1}{T},
two orders of magnitude better than that in Refs.~\cite{USMChA1,USMChA2},
and sufficient to observe the predicted aMChA in $\alpha $-quartz. aMChA
measurements were performed at room temperature on several samples cut from
natural $\alpha $-quartz crystals, the handedness of which was determined
from the sign of their optical rotation.

\begin{figure}[t]
\centering  \includegraphics[width=\linewidth]{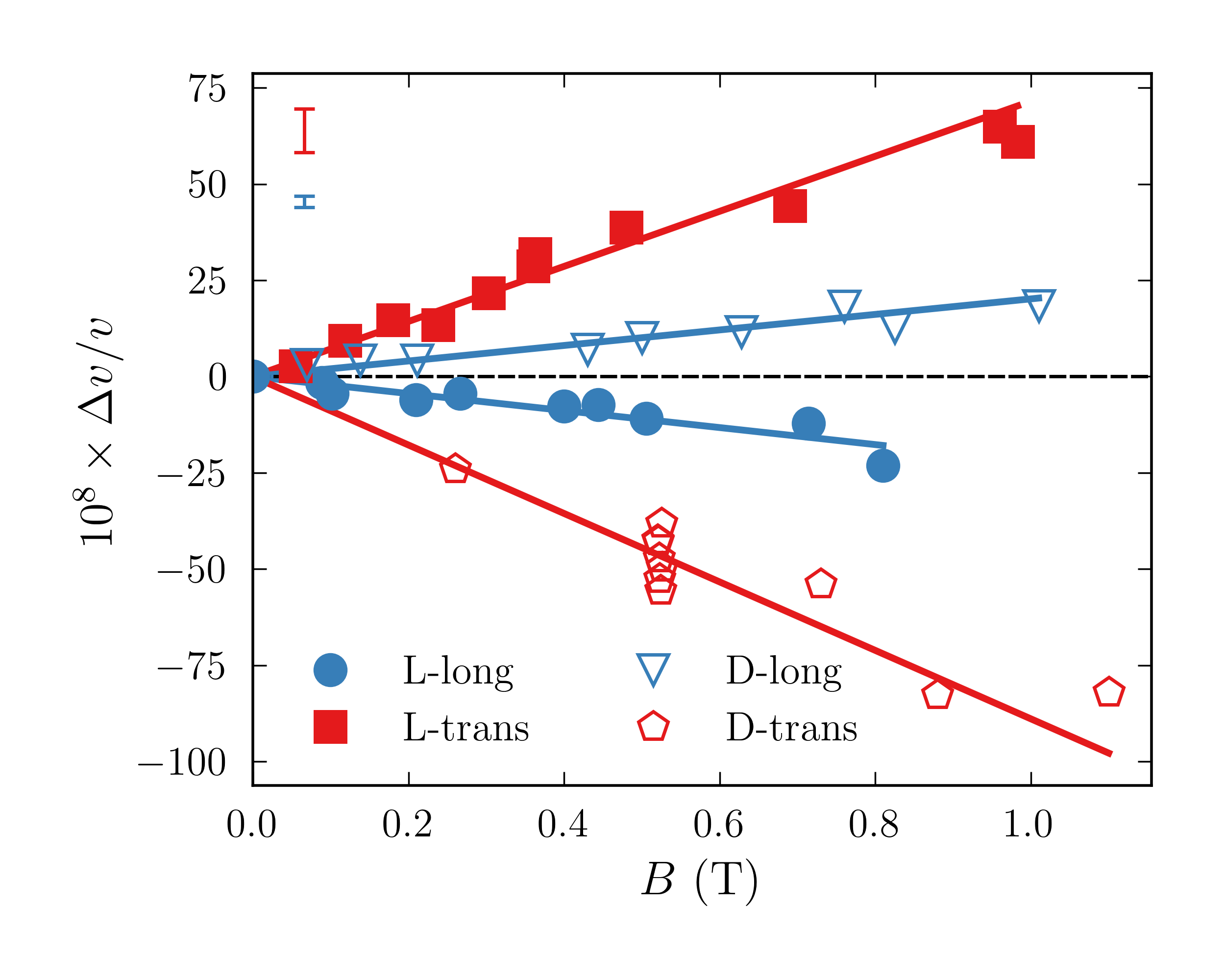} 
\caption{aMChA for longitudinal (blue) and transverse (red) waves in L
(full) and D (open) $\protect\alpha$-quartz crystals. Symbols: circles
(L-long, 481~MHz), squares (L-trans, 352~MHz), triangles (D-long, 340~MHz),
and pentagons (D-trans, 325~MHz). Solid lines are linear fits. Vertical bars
indicate typical uncertainties.}
\label{fig:dv_over_v_vs_b}
\end{figure}

Figure~\ref{fig:dv_over_v_vs_b} shows the results for $\Delta v/v_{0}$ as a
function of the magnetic field amplitude, for longitudinal and linearly
polarized transverse waves, obtained for a $D$ and an $L$ crystal, clearly
showing a linear magnetic-field dependence and opposite signs for the two
handednesses, both defining characteristics of MChA. Figure~\ref%
{fig:dv_over_v_vs_f} shows the results for $\left|\Delta v/v_{0}\right|/B$
as a function of the ultrasound frequency, for longitudinal and transverse
waves respectively. The linear frequency dependence predicted by our simple
model is clearly confirmed.

\begin{figure}[tbp]
\centering  \includegraphics[width=\linewidth]{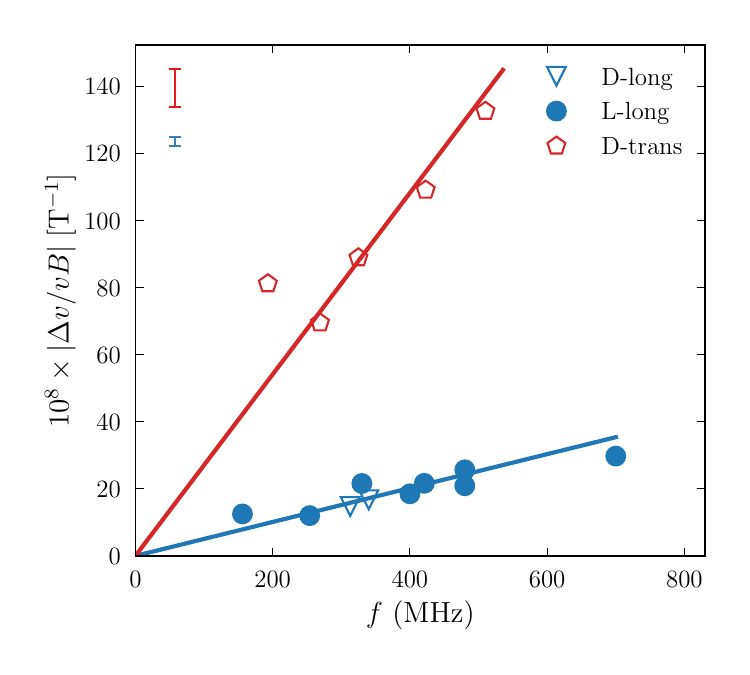}
\caption{Frequency dependence of the aMChA field slope for longitudinal
(blue) and transverse (red) waves in L (full) and D (open) $\protect\alpha$%
-quartz crystals. Symbols: circles (L-long), triangles (D-long), and
pentagons (D-trans). Solid lines are linear fits to the longitudinal and
transverse datasets. Vertical bars indicate typical uncertainties. }
\label{fig:dv_over_v_vs_f}
\end{figure}

The experimentally observed slope for longitudinal waves is $(20\pm3)\times
10^{-8}\ \mathrm{T}^{-1}$ at \SI{340}{MHz}. The prediction for this mode at
this frequency is $(8\pm3)\times 10^{-8}\ \mathrm{T}^{-1}$. The
experimentally observed slope for transverse waves at \SI{352}{MHz} is $%
(72\pm8)\times 10^{-8}\ \mathrm{T}^{-1}$. The prediction for this mode at
this frequency is $(23\pm10)\times 10^{-8}\ \mathrm{T}^{-1}$. In view of the
approximations in the Becquerel-like model, the agreement of its predictions
with experiment is quite satisfactory. The only experimental observation
that is not reproduced by the model is the opposite sign of aMChA for
longitudinal and transverse waves for a given crystal handedness. This
discrepancy may result from our assumption that Eq.~\eqref{deltav/v},
derived for transverse waves, is also valid for longitudinal waves. Better
agreement can be expected from detailed lattice dynamics calculations
including an external magnetic field, something that does not yet seem to
exist.

The equivalent measurement of oMChA has also been reported for $\alpha$%
-quartz, yielding $(\Delta v/v_{0})/B = 6\times 10^{-9}\ \mathrm{T}^{-1}$ at
an optical wavelength of \SI{632}{nm} \cite{Kalugin}, i.e. three orders of
magnitude weaker than aMChA at the same wavelength.

As no specific characteristics of $\alpha$-quartz, other than the values of
some general parameters, enter our model, one can safely assume that all
diamagnetic chiral crystals will possess aMChA. This in turn implies that
also the thermal conductivity of all dielectric chiral crystals will show
MChA, at least at low temperatures where acoustic phonons dominate the heat
transport. Our model provides guidelines for what parameters of chiral
diamagnetic crystals could lead to significant aMChA. Unfortunately, only
very few experimental data exist on the dispersion of acoustic activity or
on the acoustic Faraday effect in diamagnetic crystals, and, to our
knowledge, none are currently available on other chiral diamagnetic crystals.

A special case consists of diamagnetic crystals of chiral molecules, where
the lattice building blocks themselves can also have electronic and
vibrational chirality, which can couple to ultrasound propagation. It is
unclear whether our macroscopic model can provide a good description of this
case. The observation of aMChA for longitudinal waves in crystals suggests
that aMChA could also exist for sound propagation in chiral gases and
liquids, and in their thermal conductivity. Our model does most likely not
apply to such media. It will also not apply to paramagnetic crystals, but we
do expect a significant aMChA to exist for those too. The observation of
aMChA in diamagnetic crystals implies also the existence of the inverse
effect in such crystals, the generation of a longitudinal magnetization when
unpolarized acoustic waves, or a heat flux, traverse a chiral crystal, the
sign of which depends on the handedness of the crystal.

In conclusion, we have observed for the first time acoustic magneto-chiral
anisotropy in a diamagnetic crystal, and have developed a simple
macroscopic, Becquerel-type, model to describe our observations. This result
opens a huge class of materials (see e.g. \cite{Chiral-phonon-database} for
inorganic chiral crystals and \cite{Rekis} for molecular chiral crystals)
for the study of aMChA and its possible applications in phononics.

\textbf{Acknowledgements} This study has been partially supported through
the EUR grant NanoX n$^\circ$ ANR-17-EURE-0009 in the framework of the
``Programme des Investissements d'Avenir'' and the IRL ``Fronti\`{e}res
quantiques''.


\begin{thebibliography}{0}%
\makeatletter
\providecommand \@ifxundefined [1]{%
 \@ifx{#1\undefined}
}%
\providecommand \@ifnum [1]{%
 \ifnum #1\expandafter \@firstoftwo
 \else \expandafter \@secondoftwo
 \fi
}%
\providecommand \@ifx [1]{%
 \ifx #1\expandafter \@firstoftwo
 \else \expandafter \@secondoftwo
 \fi
}%
\providecommand \natexlab [1]{#1}%
\providecommand \enquote  [1]{``#1''}%
\providecommand \bibnamefont  [1]{#1}%
\providecommand \bibfnamefont [1]{#1}%
\providecommand \citenamefont [1]{#1}%
\providecommand \href@noop [0]{\@secondoftwo}%
\providecommand \href [0]{\begingroup \@sanitize@url \@href}%
\providecommand \@href[1]{\@@startlink{#1}\@@href}%
\providecommand \@@href[1]{\endgroup#1\@@endlink}%
\providecommand \@sanitize@url [0]{\catcode `\\12\catcode `\$12\catcode `\&12\catcode `\#12\catcode `\^12\catcode `\_12\catcode `\%12\relax}%
\providecommand \@@startlink[1]{}%
\providecommand \@@endlink[0]{}%
\providecommand \url  [0]{\begingroup\@sanitize@url \@url }%
\providecommand \@url [1]{\endgroup\@href {#1}{\urlprefix }}%
\providecommand \urlprefix  [0]{URL }%
\providecommand \Eprint [0]{\href }%
\providecommand \doibase [0]{https://doi.org/}%
\providecommand \selectlanguage [0]{\@gobble}%
\providecommand \bibinfo  [0]{\@secondoftwo}%
\providecommand \bibfield  [0]{\@secondoftwo}%
\providecommand \translation [1]{[#1]}%
\providecommand \BibitemOpen [0]{}%
\providecommand \bibitemStop [0]{}%
\providecommand \bibitemNoStop [0]{.\EOS\space}%
\providecommand \EOS [0]{\spacefactor3000\relax}%
\providecommand \BibitemShut  [1]{\csname bibitem#1\endcsname}%
\let\auto@bib@innerbib\@empty
\end{thebibliography}%


\begin{thebibliography}{99}
\bibitem{WagniereMeier} G. Wagni\`{e}re and A. Meier, Chem. Phys. Lett. 
\textbf{93}, 78 (1982).

\bibitem{BarronVbrancich} L. Barron and J. Vbrancich, Mol. Phys. \textbf{51}%
, 715 (1984).

\bibitem{Naturemca} G.L.J.A. Rikken and E. Raupach, Nature \textbf{390}, 493
(1997).

\bibitem{kleindienst} P. Kleindienst and G. Wagni\`{e}re, Chem. Phys. Lett. 
\textbf{288}, 89 (1998).

\bibitem{mcaabs} G.L.J.A. Rikken and E. Raupach, Phys. Rev. \textbf{E 58},
5081 (1998).

\bibitem{MicrowaveMChA} S. Tomita, K. Sawada, A. Porokhnyuk, and T. Ueda,
Phys. Rev. Lett. \textbf{113}, 235501 (2014). Y. Okamura, F. Kagawa, S.
Seki, M. Kubota, M. Kawasaki, and Y. Tokura, Phys. Rev. Lett. \textbf{114},
197202 (2015).

\bibitem{XrayMChA} M. Ceol\'{\i}n, S. Goberna-Ferr\'{o}n and J. R. Gal\'{a}%
n-Mascar\'{o}s, Adv. Mater. \textbf{24}, 3220 (2012); R. Sessoli, M. Boulon,
A. Caneschi, M. Mannini, L. Poggini, F. Wilhelm and A. Rogalev, Nat. Phys. 
\textbf{11}, 69 (2015).

\bibitem{emchaprl} G.L.J.A. Rikken, J. F\"{o}lling and P. Wyder, Phys. Rev.
Lett. \textbf{87}, 236602 (2001).

\bibitem{eMChACNT} V. Krsti\'{c}, S. Roth, M. Burghard, K. Kern and G.L.J.A.
Rikken, J. Chem. Phys. \textbf{117}, 11315 (2002).

\bibitem{Pop} F. Pop, P. Auban-Senzier, E. Canadell, G. L. J. A. Rikken and
N. Avarvari, Nat. Commun. \textbf{5}, 3757 (2014).

\bibitem{Yokouchi} T. Yokouchi \textit{et al.}, Nat. Commun. \textbf{8}, 866
(2017).

\bibitem{Aoki} R. Aoki, Y. Kousaka and Y. Togawa, Phys. Rev. Lett. \textbf{%
122}, 057206 (2019).

\bibitem{MChAsupercon} F. Qin \textit{et al.}, Nat. Commun. \textbf{8},
14465 (2017); R. Wakatsuki \textit{et al.}, Sci. Adv. \textbf{3}, e1602390
(2017).

\bibitem{RikkenTe} G.L.J.A. Rikken and N. Avarvari, Phys. Rev. B \textbf{99}%
, 245153 (2019).

\bibitem{MagnonMChA} S. Seki \textit{et al.}, Phys. Rev. B \textbf{93},
235131 (2016).

\bibitem{USMChA1} T. Nomura, X.-X. Zhang, S. Zherlitsyn, J. Wosnitza, Y.
Tokura, N. Nagaosa and S. Seki, Phys. Rev. Lett. \textbf{122}, 145901 (2019).

\bibitem{USMChA2} T. Nomura \textit{et al.}, Phys. Rev. Lett. \textbf{130},
176301 (2023).

\bibitem{dMChA-Rikken} G.L.J.A. Rikken and N. Avarvari, Nat. Commun. \textbf{%
13}, 3564 (2022).

\bibitem{Theory-ioMChA} G. Wagni\`{e}re, Phys. Rev. \textbf{A40}, 2437
(1989).

\bibitem{Theory-ieMChA} T. Yoda, T. Yokoyama and S. Murakami, Sci. Rep. 
\textbf{5}, 12024 (2015).

\bibitem{Exp-ieMChA} T. Furukawa, Y. Shimokawa, K. Kobayashi and T. Itou,
Nat. Commun. \textbf{8}, 954 (2017).

\bibitem{Non-reciproc-review} N. Nagaosa and Y. Yanase, Annu. Rev. Condens.
Matter Phys. \textbf{15}, 63 (2024).

\bibitem{Review-MChA} M. Atzori, C. Train, E. A. Hillard, N. Avarvari and
G.L.J.A. Rikken, Chirality \textbf{33}, 844 (2021).

\bibitem{ValidationMChA} M. Atzori \textit{et al.}, Sci. Adv. \textbf{7},
eabg2859 (2021).

\bibitem{Phonon-MChA-Weyl} S. Sengupta, M. Nabil, Y. Lhachemi and I. Garate,
Phys. Rev. Lett. \textbf{125}, 146402 (2020).

\bibitem{Sup-mat} Supplementary material.

\bibitem{Portigal-Burstein} D. L. Portigal and E. L. Burstein, Phys. Rev. 
\textbf{170}, 673 (1968).

\bibitem{Shen} Z.-G. Shen \textit{et al.}, Phys. Rev. B \textbf{45}, 12834
(1992).


\bibitem{Pine} A. S. Pine, Phys. Rev. B \textbf{2}, 2049 (1970).

\bibitem{Joffrin} J. Joffrin and A. Levelut, Solid State Commun. \textbf{8},
1573 (1970).

\bibitem{Bialas} H. Bialas and G. Schauer, Phys. Status Solidi A \textbf{72}%
, 679 (1982).

\bibitem{Joffrin1980} C.~Joffrin, B.~Dorner and J.~Joffrin, J. Physique
Lett. \textbf{41}, 391 (1980).

\bibitem{Chiral-phonon-database} Chiral phonon materials database, %
\url{https://materialsfingerprint.com/}

\bibitem{Hao} H.-Y. Hao and H. J. Maris, Phys. Rev. B \textbf{63}, 224301
(2001).

\bibitem{Kalugin} N. G. Kalugin, P. Kleindienst and G. H. Wagni\`{e}re,
Chem. Phys. \textbf{248}, 105 (1999).

\bibitem{Rekis} T. Rekis, Acta Crystallogr. B \textbf{76}, 307 (2020).
\end{thebibliography}
\end{document}